\documentstyle[epsfig,url,times]{aa}

\def\B{{\em BeppoSAX}}                       
\def\O{OAO1657--415}
\def\Ec{\mbox{$E_{\rm cyc}$}}
\def\mcc#1{\multicolumn{2}{c}{#1}}

\title{The broad-band spectrum of \O\ with \B: In search of cyclotron lines}

\author{M.~Orlandini\inst{1}
\and D.~Dal~Fiume\inst{1}
\and S.~Del~Sordo\inst{2}
\and F.~Frontera\inst{1,3}
\and A.N.~Parmar\inst{4}
\and A.~Santangelo\inst{2}
\and A.~Segreto\inst{2}
}

\institute{
Istituto Tecnologie e Studio Radiazioni Extraterrestri, TeSRE/CNR, 
 via Gobetti 101, 40129 Bologna, Italy
\and
Istituto Fisica Cosmica e Applicazioni all'Informatica, IFCAI/CNR,
 via La~Malfa 153, 90146 Palermo, Italy
\and
Dipartimento di Fisica, Universit\`a di Ferrara, 
 via Paradiso 11, 44100 Ferrara, Italy
\and Astrophysics Division, Space Science Department of ESA, ESTEC,
 NL--2200 AG Noordwijk, The Netherlands
}

\offprints{orlandini@tesre.bo.cnr.it}
\date{Received \today; Accepted }

\thesaurus{06 (         
               02.13.1  
               08.02.1  
               08.14.1  
               08.16.7  
               13.25.5  
              )
}

\begin{document}


\maketitle\markboth{M.~Orlandini et al.}{\B\ observation of \O}

\begin{abstract}

We report on a 30~ks observation of the high-mass X--ray binary pulsar
\object{\O} performed by \B\ on September 1998. The wide band spectrum is well
fit by both a cutoff power law, or a power law modified by a high energy
cutoff, plus a fluorescence Iron line at 6.5~keV. The two models are
statistically equivalent. The inclusion of a cyclotron resonance feature at
$\sim$36~keV -- corresponding to a magnetic field strength $3.2\,(1+z)\times
10^{12}$~G, where $z$ is the gravitational redshift -- improves significantly
the $\chi^2$ for the cutoff model but only marginally for the power law plus
high energy cutoff model. The statistical significance of our data is not
adequate to discriminate between the two models, and even the {\em Normalized
Crab Ratio} technique, successfully used to pinpoint cyclotron features in the
spectra of other X--ray pulsars, is not conclusive in answering the question
whether the feature is real or it is an artifact due to an improper modeling of
the continuum used to fit the data.


\end{abstract}

\section{Introduction}

X--ray binary pulsars (XRBs) display a common signature in their X--ray
spectra: the presence of a high-energy cutoff that makes their spectra rapidly
drop above $\sim$40--50~keV. There are only few exceptions to this general
behaviour, and \O\ is one of them: indeed, this source displays one of the
hardest X--ray spectra amongst XRBs, with no steepening up to 100~keV
(\cite{683}). In those cases in which cyclotron line features are observed, the
cutoff energy $E_{\rm c}$ seems to be related to the cyclotron line energy \Ec\
through a phenomenological relation found by \nocite{407} Makishima \& Mihara
(1992): $\Ec\simeq (1.2-2.5)\cdot E_{\rm c}$. Because \Ec\ is related to the
neutron star magnetic field strength $B_{12}$ in units of $10^{12}$~G by the
relation $\Ec = 11.6\,B_{12}\,(1+z)^{-1}$~G, where $z$ is the gravitational
redshift, a measurement of $E_{\rm c}$ can give an estimate of the neutron star
magnetic field. For \O\ we therefore expect $B_{12}\ga 10$: only another XRB,
\object{A0535+26}, is known to possess such a strong magnetic field
(\cite{375}).

\O\ was discovered by the {\em Copernicus} satellite (\cite{1659}), and soon
after its discovery a 38~s pulsation was detected by the A2 experiment aboard
{\em HEAO~1} (\cite{684}). The optical counterpart of \O\ was first identified
with the spectroscopic binary \object{V861 Sco}, but subsequent observations
(\cite{710}; \cite{711}) produced positional error boxes that excluded this
star as the optical counterpart that is, up today, still unknown. Since 1991
\O\ has been monitored continuously by the BATSE experiment aboard {\em
CGRO\/}. This allowed the discovery of X--ray eclipses by the stellar companion
and the determination of a 10.44~d orbital period (\cite{218}). The distance to
the source is not known, but a lower limit of 11~kpc has been determined by
assuming that the neutron star is rotating at its equilibrium period. Because
of its low galactic latitude, the low energy ($\la$2~keV) spectrum of \O\ is
very absorbed, with N$_{\rm H}\ga 10^{23}$~cm$^{-2}$. {\em GINGA} observations
of the source revealed the presence of a soft excess, below 3~keV, modeled with
a 0.3~keV blackbody (\cite{1811}). The \O\ X--ray spectrum also displays a
fluorescence Iron line at $\sim$6.6~keV (\cite{684}; \cite{1811}).

\begin{figure*}
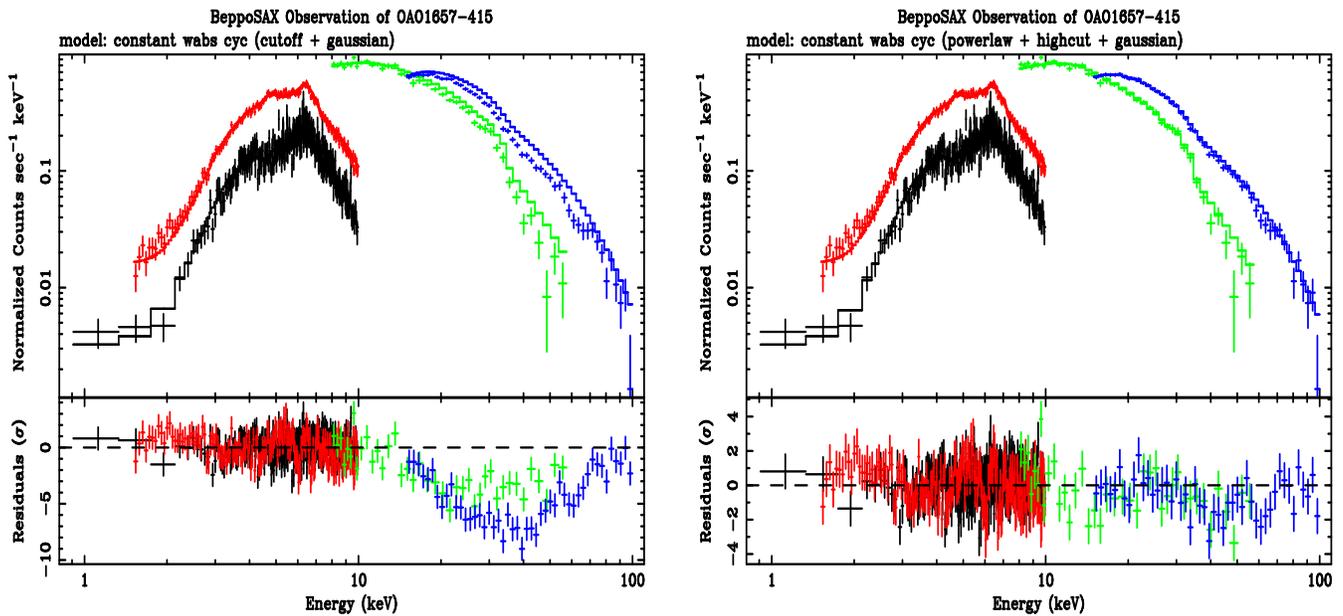

\centerline{%
\epsfig{file=Bf081.f1a,width=0.45\textwidth,height=8.5cm,angle=270}\qquad
\epsfig{file=Bf081.f1b,width=0.45\textwidth,height=8.5cm,angle=270}}
\caption[]{\O\ count rate spectrum {\em (plus signs)} and best-fit continuum
model {\em (histogram)\/}, together with the fit residuals. On the left the EC
model was used, on the right the HEC model. The CRF normalization has been set
to zero, in order to show its profile in the residual panel}
\label{fig:fits}
\end{figure*}

\section{X--ray observation and data analysis}

\O\ was observed by the Narrow Field Instruments (NFIs) aboard the
Italian-Dutch X--ray satellite \B\ (\cite{1530}) from 1998 September 4 15:41:46
to September 5 07:20:35 UT. All the NFIs performed nominally during the
observation, namely the two imaging instruments LECS (0.1--10~keV; \cite{1531})
and MECS (1.5--10~keV; \cite{1532}), and the two mechanically collimated
instruments HPGSPC (4--180~keV; \cite{1533}) and PDS (15--200~keV;
\cite{1386}). Data were collected in direct mode, providing information on
time, energy, and position. Good data were selected using default criteria. The
net exposure times for the four NFIs were 9.2, 26, 11.7, and 10.8~ks for the
LECS, MECS, HPGSPC, and PDS, respectively. These differences are due to the
constraint that the LECS must operate in satellite night time, and rocking of
the collimated NFIs. For the imaging NFIs data were extracted from $4'$
circular regions centered on the source position.

The background for the imaging NFIs was estimated from standard background
blank field measurements. Because of its low galactic latitude, and therefore
possible contamination by background sources, we also extracted spectra from
two blank $4'$ circular regions offset by $11'$ with respect to the source. Our
result on the spectral analysis does not depend on which background was used:
for example, the 6--7~keV MECS count rate in the offset backgrounds is less
than 1\% of the source count rate in the same band. For the non-imaging NFIs
the background was monitored by rocking the collimators off-source by $3^\circ$
for HPGSPC (one-side rocking), and $3\fdg 5$ for PDS (two-side rocking). The
off-source fields were checked for the presence of contaminating sources: one
of the PDS background fields was found to show an excess count rate with
respect to the other, a clear signature of contaminating sources, and therefore
was not used for the estimation of the source background.

The contribution of the Galactic Ridge (GR) emission was estimated by using the
results from {\em RXTE} observations of the galactic plane (\cite{1729}). We
found that the ratio between the GR and the source flux is always $\la
10^{-3}$, therefore we exclude any contribution of the GR emission to our
spectral data.

The spectra of each \B\ NFI were first fit separately to derive the best model
able to describe the broad band spectrum. We find that the LECS and MECS
spectra can be described in terms of an absorbed power law plus a Gaussian line
in emission, while the PDS spectrum clearly presents a deviation from a pure
power law, with a steepening above $\sim$20~keV. We therefore used a power law,
modified at low energy by photoelectrical absorption and at high energy by a
cutoff, to describe the broad band spectrum. Variable normalizations among the
NFIs, found in agreement with those obtained from the fit to the Crab spectrum,
were introduced to take into account known calibration uncertainties. Both an
exponential cutoff (EC) and a high-energy cutoff (HEC; \cite{303}) yield a good
fit, while a thermal bremsstrahlung did not fit our data.  Also the smoother
Fermi-Dirac cutoff (\cite{1584}) did not adequately describe the high energy
tail. An F-test shows that the probability of chance improvement (PCI) of the
$\chi^2$ between the EC and HEC models is 78\%, therefore they are
statistically equivalent. In Table~\ref{tab:best_fits} we present the best-fit
parameters for both the two models.

Concerning the low ($\la$3~keV) energy part of the spectrum, our data do not
show any soft excess as observed by {\em GINGA} (\cite{1811}). The Iron line
equivalent width is $360\pm 280$~eV, in agreement with the {\em GINGA} result
but lower than the {\em HEAO~1} measurement (\cite{684}). This confirms the
possible contamination by the GR emission of the latter, as discussed by Kamata
et~al.\ (1990). The Iron line energy found by \B\ is lower than that measured
by {\em GINGA} ($6.60\pm 0.08$~keV) and {\em HEAO~1} ($6.7\pm 0.1$~keV). The
unabsorbed fluxes in the bands 2--10 and 10--100~keV are $4.7\times 10^{-10}$,
and $1.8\times 10^{-9}$~erg~cm$^{-2}$~s$^{-1}$, respectively. The total
0.1--100~keV X-ray luminosity is $2.5\times 10^{35}$ erg~s$^{-1}$~d$^2_{\rm
kpc}$, where d$_{\rm kpc}$ is the \O\ distance in kpc.

\begin{table*}
\begin{flushleft}
\caption[]{Best fit parameters for the two models that fit the \O\ continuum: a
cutoff power law and a power law plus high energy cutoff. For both the models
we show the best fit parameters without the cyclotron resonance feature (CRF)
and with it. Because the CRF width is not well constrained by the fit, we
present also the best-fit parameters for a CRF width fixed at 10~keV}
\label{tab:best_fits}
\begin{tabular}{l*{5}{r@{$\,\pm\,$}l}}
\\ \hline\noalign{\smallskip}
Parameter       & \multicolumn{10}{c}{Value$^a$} \\ \noalign{\smallskip} \cline{2-11} \noalign{\smallskip}
                & \multicolumn{6}{c}{Cutoff} & \multicolumn{4}{c}{Power Law + Highcut} \\
                & \multicolumn{2}{c}{w/o CRF} & \multicolumn{2}{c}{w/ CRF} & 
		  \multicolumn{2}{c}{w/ CRF} &
                  \multicolumn{2}{c}{w/o CRF} & \multicolumn{2}{c}{w/ CRF} \\
\noalign{\smallskip} \hline \noalign{\smallskip}
N$_{\rm H}$ ($10^{22}$ cm$^{-2}$) & 12.1 & 0.4                   & 12.0 & 0.4                   & 12.1 & 0.4                   & \mcc{$12.7_{-0.2}^{+0.4}$}   & \mcc{$12.7_{-0.4}^{+0.3}$}    \\
$\alpha$                          & 0.41 & 0.04                  & \mcc{$0.44_{-0.06}^{+0.11}$} & 0.43 & 0.04                  & 0.83 & 0.04                  & \mcc{$0.83_{-0.06}^{+0.03}$}  \\
E$_{\rm c}$ (keV)                 & 16.4 & 0.5                   & \mcc{$19_{-2}^{+18}$}        & 17.5 & 0.7                   & \mcc{$13.0_{-0.4}^{+0.5}$}   & \mcc{$12.9_{-0.4}^{+0.5}$}    \\
E$_{\rm f}$ (keV)                 & \mcc{}                       & \mcc{}                       & \mcc{}                       & \mcc{$21.2_{-0.4}^{+0.6}$}   & \mcc{$22.2_{-0.7}^{+0.8}$}    \\
E$_{\rm Fe}$ (keV)                & 6.48 & 0.04                  & 6.48 & 0.04                  & 6.48 & 0.04                  & 6.47 & 0.04                  & 6.47 & 0.04                   \\
$\sigma_{\rm Fe}$ (keV)           & \mcc{$0.33_{-0.07}^{+0.08}$} & \mcc{$0.35_{-0.07}^{+0.08}$} & \mcc{$0.34_{-0.07}^{+0.08}$} & 0.42 & 0.05                  & \mcc{$0.42_{-0.06}^{+0.05}$}  \\
E$_{\rm CRF}$ (keV)               & \mcc{}                       & \mcc{$30_{-25}^{+7}$}        & 36 & 3                       & \mcc{}                       & \mcc{$44_{-5}^{+11}$}         \\
$\tau_{\rm CRF}$                  & \mcc{}                       & \mcc{$0.18_{-0.13}^{+0.09}$} & 0.12 & 0.05                  & \mcc{}                       & \mcc{$0.11_{-0.04}^{+0.06}$}  \\
W$_{\rm CRF}$ (keV)               & \mcc{}                       & \mcc{$28_{-15}^{+26}$}       & \mcc{10 (fixed)}             & \mcc{}                       & \mcc{$11_{-7}^{+19}$}         \\
$\chi^2_\nu$ (dof)                & \mcc{1.129 (572)}            & \mcc{1.088 (569)}            & \mcc{1.094 (570)}            & \mcc{1.127 (571)}            & \mcc{1.117 (568)} \\
\noalign{\smallskip} \hline \noalign{\smallskip}
\multicolumn{7}{l}{$^a$ Uncertainties at 90\% confidence for a single parameter}
\end{tabular}
\end{flushleft}
\end{table*}

From the form of the residuals above 20~keV we were led to add an extra
component to our spectral model, that we interpret in terms of a cyclotron
resonance feature (CRF). Both a Lorenzian (\cite{1547}) and a Gaussian in
absorption (\cite{1172}) improved the fit, and here we will present results
obtained with the Lorenzian form of the CRF. The inclusion of the extra
component have different significance for the two continuum models. Indeed, an
F-test shows that the inclusion of the CRF improves significantly the $\chi^2$
for the EC model (the PCI is $2.8\times 10^{-5}$), while the PCI for the HEC
model is higher ($1.1\times 10^{-2}$). In Table~\ref{tab:best_fits} we show the
best-fit parameters of the CRF, and in Fig.~\ref{fig:fits} we show both the EC
and HEC best-fit spectral models, with the residual CRF Lorenzian line profile
shown in the residual panel, obtained by setting the line normalization to
zero. We can see very well how the line is significant and very broad by using
the EC model, and with much less significance with the HEC model. Because the
line width is not well constrained by the fit, we also present in
Table~\ref{tab:best_fits} the best-fit parameters obtained by fixing the CRF
width at 10~keV.

\begin{figure}
\epsfig{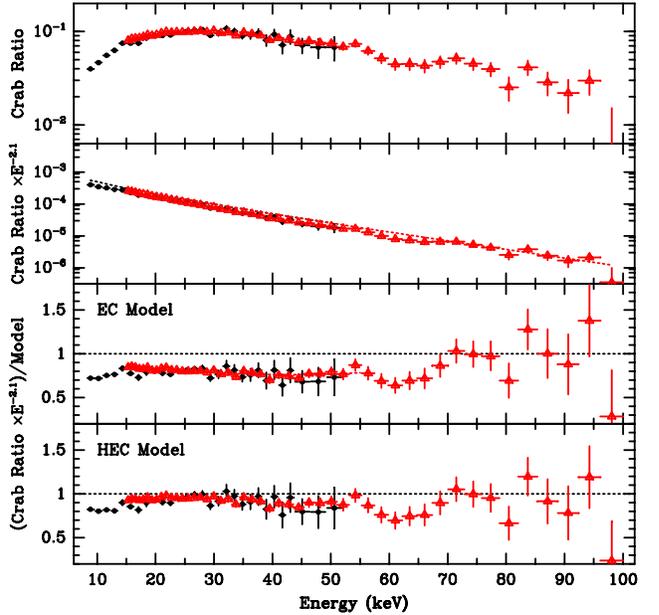}
\vspace{-1cm}
\caption[]{\O\ Normalized Crab Ratio. {\em Top\/:} Ratio between the HPGSPC
{\em (black dots)} and PDS {\em (grey triangles)} count rate spectra with the
Crab pulsar one. {\em Second Panel\/}: Crab ratio multiplied by the functional
form of the Crab spectrum, a featureless power law of spectral index 2.1. {\em
Third Panel\/}: Crab ratio times $E^{-2.1}$ divided by the EC continuum. {\em
Fourth Panel \/}: The same as the third panel but divided by the HEC continuum}
\label{fig:crab_ratio}
\end{figure}

In order to check if the CRF is real or an artifact due to an improper modeling
of the continuum, we calculated the ratio between the HPGSPC and PDS count rate
spectra with the Crab pulsar count spectrum. This ratio has the advantage of
minimizing calibration uncertainties. In Fig.~\ref{fig:crab_ratio}, top panel,
we can see that this ratio does not show any evident change of slope, as
observed in other pulsars (see e.g.\ \cite{1622}; \cite{1583}). By multiplying
the Crab ratio by the functional form of the Crab spectrum
(Fig.~\ref{fig:crab_ratio}, second panel), {\em i.e.\/} a power law of index
2.1, and diving by the functional form of the continuum
(Fig.~\ref{fig:crab_ratio}, third and fourth panels), as obtained from our
spectral fits, we are able to emphasize any deviations from the continuum. This
{\em Normalized Crab Ratio} technique has been successfully used to pinpoint
CRFs in the spectra of other X--ray pulsars (\cite{1946}). The third panel of
Fig.~\ref{fig:crab_ratio} shows a broad line, modeled with a Gaussian line
centered at $43_{-5}^{+18}$~keV and $\sigma_{\rm cyc} = 28_{-6}^{+17}$~keV
($\chi^2/{\rm dof} = 17.7/30$).

The narrow feature at $\sim$60~keV present in the PDS Crab ratio has $\sigma =
5_{-2}^{+5}$~keV if modeled with a Gaussian in 53--70~keV ($\chi^2/{\rm dof} =
0.18/4$). This value is marginally comparable with the PDS energy resolution of
8.6~keV (FWHM; \cite{1386}) at $62_{-2}^{+3}$~keV. The inclusion of this
feature in the broad-band fit yields a $\chi^2/{\rm dof}$ of 618.19/567. An
F-test shows that the PCI of the $\chi^2$ is 17.5\%. For these reasons we do
not consider it as a real feature.

\section{Discussion}

The main result obtained from the \B\ observation of \O\ is the clear detection
of a steepening of the spectrum above 20~keV, and that this drop occurs in two
steps. Indeed \B\ observations of XRBs have shown that there are actually {\em
two} changes of slope in the high (E$\ga$10~keV) energy part of their spectra:
a first change of slope occurs in the $\sim$10--20~keV range, while a second
steepening occurs for higher energies. This is partly expressed in the form of
the high-energy cutoff (\cite{303}), that has been used to empirically describe
the two-step spectral drop in these sources. In the XRB Vela X--1, for example,
the first change of slope occurs at $\sim$24~keV and it can be misinterpreted
as a CRF (\cite{1581}). For \O\ we think we are observing the same phenomenon:
the high energy part of the spectrum presents two changes of slope: if we model
the continuum with only one steepening, then the fit requires the presence of a
second one, that can be modeled with a CRF. On the other hand, if we model the
continuum with a two-step change of slope (the HEC model) then an additional
CRF is not required to describe the spectral drop. The fact that we observe a
broad CRF is due to the fact that the \O\ spectrum is very hard, therefore the
steepening is smooth. On the other hand, in the case of a softer spectrum, like
in Vela X--1, the steepening is much more abrupt, and therefore the {\em false}
CRF is narrower.

\begin{figure}
\epsfig{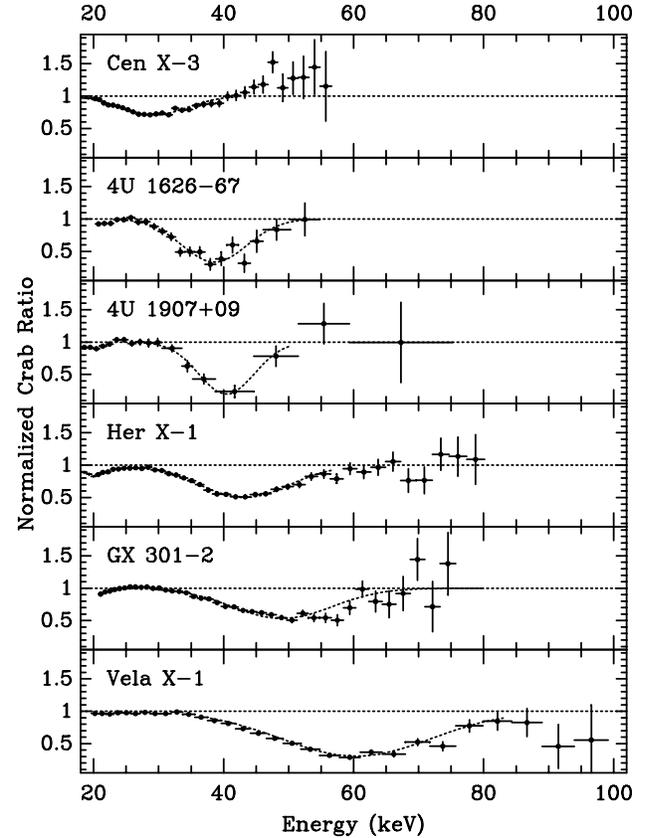}
\vspace{0.7cm}
\caption[]{Normalized Crab Ratios from PDS data obtained for a sample of XRBs
observed by \B\ (see e.g.\ \cite{1946}). The continuum drop is modeled with a
HEC for all the sources but Cen X--3, for which an EC model is used}
\label{fig:crab_ratios_all}
\end{figure}

Lacking a model of the continuum able to describe this behaviour, the only way
to discriminate a real CRF from an artifact due to an improper modeling of the
continuum is through the Crab ratio, which is model independent. In
Fig.~\ref{fig:crab_ratios_all} we show the Normalized Crab Ratios from PDS data
obtained for a sample of XRBs observed by \B. A comparison with
Fig.~\ref{fig:crab_ratio} clearly shows that the \O\ Crab ratio does not
pinpoint the CRF as it does for all the other XRBs. The fact that \Ec\ depends
on the adopted continuum seems to rule against a {\em real} CRF, but it is
intriguing that the Makishima \& Mihara relation is satisfied for the \O\ CRF.
Furthermore, the relation between \Ec\ and its FWHM, clearly shown in
Fig.~\ref{fig:crab_ratios_all} where the CRF width increases as \Ec\ increases,
and explained in terms of Doppler broadening of the electrons responsible for
the resonance (\cite{1614}) is fulfilled if we assume for the CRF width a value
of 10~keV, which is the one obtained from the HEC fit, and can be considered as
a lower limit for the width. In this case the \O\ surface magnetic field
strength would be ($3.1\pm 0.2)\times 10^{12}\, (1+z)$~G. On the other hands,
this relation would not be satisfied by the broader CRF. Pulse phase
spectroscopy should be able to resolve this ambiguity. However, longer
observations may be needed.

\begin{acknowledgements}
This research is supported in part by the Italian Space Agency.
\end{acknowledgements}


\begin{thebibliography}{}

\bibitem[Armstrong et~al.\ 1980]{710}
Armstrong J.T., Johnston M., Bradt H.V., et~al., 1980,
\newblock ApJ, 236, L131

\bibitem[Boella et~al.\ 1997a]{1530}
Boella G., Butler R.C., Perola G.C., et~al., 1997a,
\newblock A\&AS, 122, 299

\bibitem[Boella et~al.\ 1997b]{1532}
Boella G., Chiappetti L., Conti G., et~al., 1997b,
\newblock A\&AS, 122, 327

\bibitem[Byrne et~al.\ 1981]{683}
Byrne P.F., Levine A.M., Bautz M., et~al., 1981,
\newblock ApJ, 246, 951

\bibitem[Chakrabarty et~al.\ 1993]{218}
Chakrabarty D., Grunsfeld J., Prince T.A., et~al., 1993,
\newblock ApJ, 403, L33

\bibitem[Dal~Fiume et~al.\ 1998]{1583}
Dal~Fiume D., Orlandini M., Cusumano G., et~al., 1998,
\newblock A\&A, 329, L41

\bibitem[Dal~Fiume et~al.\ 1999]{1946}
Dal~Fiume D., Orlandini M., Del~Sordo S., et~al., 1999,
\newblock ASR, in press (astro-ph/9906086)

\bibitem[Frontera et~al.\ 1997]{1386}
Frontera F., Costa E., Dal~Fiume D., et~al., 1997,
\newblock A\&AS, 122, 357

\bibitem[Grove et~al.\ 1995]{375}
Grove J.E., Strickman M., Johnson W.N., et~al., 1995,
\newblock ApJ, 438, L25

\bibitem[Kamata et~al.\ 1990]{1811}
Kamata Y., Koyama K., Tawara Y., et~al., 1990,
\newblock PASJ, 42, 785

\bibitem[Makishima \& Mihara 1992]{407}
Makishima K., Mihara T., 1992,
\newblock in Frontiers of X--ray Astronomy, eds. Tanaka Y., Koyama K.
  {Universal Academy Press, Tokyo}, 23

\bibitem[Manzo et~al.\ 1997]{1533}
Manzo G., Giarrusso S., Santangelo A., et~al., 1997,
\newblock A\&AS, 122, 341

\bibitem[M{\'e}sz{\'a}ros 1992]{1614}
M{\'e}sz{\'a}ros, P. 1992,
\newblock {High-Energy Radiation from Magnetized Neutron Stars},
\newblock Chicago Univ. Press

\bibitem[Mihara 1995]{1547}
Mihara T., 1995,
\newblock PhD thesis, RIKEN

\bibitem[Orlandini et~al.\ 1998a]{1581}
Orlandini M., Dal~Fiume D., Frontera F., et~al., 1998a,
\newblock A\&A, 332, 121

\bibitem[Orlandini et~al.\ 1998b]{1622}
Orlandini M., Dal~Fiume D., Frontera F., et~al., 1998b,
\newblock ApJ, 500, L163

\bibitem[Parmar et~al.\ 1980]{711}
Parmar A.N., Branduardi-Raymont G., Pollard G.S.G., et~al., 1980,
\newblock MNRAS, 193, 49P

\bibitem[Parmar et~al.\ 1997]{1531}
Parmar A.N., Martin D., Bavdaz M., et~al., 1997,
\newblock A\&AS, 122, 309

\bibitem[Polidan et~al.\ 1978]{1659}
Polidan R.S., Pollard G.S., Sanford P.N., et~al., 1978,
\newblock Nat, 275, 296

\bibitem[Soong et~al.\ 1990]{1172}
Soong Y., Gruber D.E., Peterson L.E., et~al., 1990,
\newblock ApJ, 348, 641

\bibitem[Tanaka 1986]{1584}
Tanaka Y., 1986,
\newblock in {Radiation Hydrodynamics in Stars and Compact Objects}, eds.
 Mihalas D., Winkler K.H. {Springer, Berlin}, 198

\bibitem[Valinia \& Marshall 1998]{1729}
Valinia A., Marshall F.E., 1998,
\newblock ApJ, 505, 134

\bibitem[White \& Pravdo 1979]{684}
White N.E., Pravdo S.H., 1979,
\newblock ApJ, 233, L121

\bibitem[White et~al.\ 1983]{303}
White N.E., Swank J.H., Holt S.S., 1983,
\newblock ApJ, 270, 711

\end{thebibliography}
\end{document}